\documentclass[]{article}

\usepackage{graphicx}
\usepackage{url , hyperref}
\usepackage{amssymb}
\usepackage{longtable} 
\usepackage{fullpage}
\usepackage{lineno}
\usepackage{float}
\usepackage{caption}
\usepackage[english]{babel} 
\usepackage{geometry} 
\usepackage{mathtools}
\renewenvironment{abstract}[1][]
 {\if\relax\detokenize{#1}\relax\else\selectlanguage{#1}\fi
  \noindent\textbf{\abstractname}\par\medskip\noindent\ignorespaces}
 {\par\bigskip}

\makeatletter
\def\ps@pprintTitle{%
 \let\@oddhead\@empty
 \let\@evenhead\@emptynto
 \def\@oddfoot{}%
 \let\@evenfoot\@oddfoot}
\makeatother

\usepackage{authblk}
\title{City limits in the age of smartphones and urban scaling\footnote{\textbf{Cite as}: Sotomayor-G\'omez B \& Samaniego H. 2020. City Limits in the Age of Smartphones and Urban Scaling. Computers, Environment and Urban Systems. 79: 101423 doi: \protect\url{http://doi.org/10.1016/j.compenvurbsys.2019.101423}}}
\author[1,2]{Boris Sotomayor-G\'omez}
\author[2,3,4]{Horacio Samaniego\footnote{Corresponding author's email: \url{horacio@ecoinformatica.cl}}}

\affil[1]{Laboratorio de Ecoinform\'atica,
    Instituto de Conservaci\'on, Biodiversidad y Territorio, 
    Facultad de Ciencias Forestales y Recursos Naturales,
    Universidad Austral de Chile, 
    Campus Isla Teja, Valdivia, 
    Chile}
\affil[2]{Instituto de Inform\'atica,
Facultad de Ciencias de la Ingenier\'ia,  Universidad Austral de Chile, 
    Campus Miraflores, Valdivia, 
    Chile}
\affil[3]{Instituto de Ecolog\'ia y Biodiversidad (IEB), Casilla 653, Santiago, Chile}
\affil[4]{Instituto de Sistemas Complejos de Valpara\'iso, Subida Artiller\'ia 470, Valpara\'iso 2360448, Chile}

\date{}                     
\providecommand{\keywords}[1]{\textbf{\textit{Keywords:}} #1}
\begin{document}







\maketitle

\begin{abstract}[English]
Urban planning still lacks appropriate standards to define city boundaries across urban systems. This issue has historically been left to administrative criteria, which can vary significantly across countries and political systems, hindering a comparative analysis across urban systems. However, the wide use of Information and Communication Technologies (ICT) has now allowed the development of new quantitative approaches to unveil how social dynamics relates to urban infrastructure. In fact, ICT provide the potential to portray more accurate descriptions of the urban systems based on the empirical analysis of millions of traces left by urbanites across the city. In this work, we apply computational techniques over a large volume of mobile phone records to define urban boundaries, through the analysis of travel patterns and the trajectory of urban dwellers in conurbations with more than 100,000 inhabitants in Chile. We created and analyzed the network of interconnected places inferred from individual travel trajectories. We then ranked each place using a spectral centrality method. This allowed to identify places of higher concurrency and functional importance for each urban environment. Urban scaling analysis is finally used as a diagnostic tool that allowed to distinguish urban from non-urban spaces. The geographic assessment of our method shows a high congruence with the current and administrative definitions of urban agglomerations in Chile. Our results can potentially be considered as a functional definition of the urban boundary. They also provide a practical implementation of urban scaling and data-driven approaches on cities as complex systems using increasingly larger non-conventional datasets.

\end{abstract}


\section{Introduction}
\label{S:1}
Urban expansion ranks as one of the most pressing issues of modern society and the consequences of continuous growth could be devastating for human welfare \cite{WUP2018}. While considerable efforts are seeking to contain urban expansion, we still lack a standardized method to measure the extent of cities boundaries. Even if we accept that the definition of urban limits is highly constrained to our understanding of urban systems, current research sets urbanites, and their interactions, as centerpieces of the urban organization and productivity. Important efforts have already contributed to the understanding of urban morphology and city dynamics \cite{barthelemy_libro}. Developing a standard and functional criteria to define the spatial boundaries of urban activity will contribute to increase our understanding of several aspects of urban life \cite{kenworthy1999patterns,echenique2001mobility,rogers2001lets,paddison2000handbook,Cranshaw2012}.

The lack of unified criteria to define urban boundaries is highlighted by the United Nations World Cities report, which indicates that countries' urban proportions face multiple non-comparable definitions \cite{unitednations,unitednations2017}. As an example, the city of Toronto in Canada, may be considered to have between 2.6 and 5.1 million inhabitants, depending on which of the three definitions of city we use to calculate its size. Such variable figures portray annual growth rates ranging from 0.9\% to 1.8\%, which may clearly lead to significant complications for efficient planning efforts.

On the other hand, important strides have been made to understand and study cities using alternative perspectives. Some of these studies proposed scaling –as observed in biology and complex systems \cite{west1997general, batty2007cities, samaniego2008cities, bon1979allometry, veregin1997allometric}– as a useful tool to understand and quantitatively predict different urban functions and characteristics \cite{bettencourt2010urban, bettencourt2016urban, van2016urban, alves2014empirical, gomez2012statistics, meirelles2018evolution}. Although several studies have recently questioned the methodology \cite{leitao2016scaling,depersin2018global}, there is little doubt about its importance \cite{barthelemy_libro,batty2013new,keuschnigg2019urban}. A candidate solution to solve this controversy lies in the unification of spatial definitions of the urban environment, as geographical delimitation has been associated with the great variability of scaling coefficients for various urban functions \cite{louf2014scaling, arcaute2015constructing, cottineau2017diverse, fragkias2013does,cottineau2019mobile}.

The search for a spatial definition of the city has not been limited to the aforementioned. The identification of an optimal methodology defining urban boundaries remains an old problem that is usually addressed by studying the spatial density distribution with the aid of comprehensive surveys and censuses. For instance, Rozenfeld et al. \cite{rozenfeld2008laws,rozenfeld2011area} centered the identification of urban systems on the notion of spatial continuity using the City Clustering Algorithm (CCA). This was later updated by Oliveira et al. to the City Local Clustering Algorithm (CLCA) improving clustering and computational capabilities with a worldwide application \cite{oliveira2018worldwide}. Along similar lines, Cranshaw et al. \cite{Cranshaw2012} proposed a spectral clustering definition of a functional neighborhood in Pittsburgh. A pioneering work by Ratti et al. \cite {ratti2010redrawing} proposed a novel approach integrating community detection algorithms built from the complex networks of over 12 billion land-line calls to map functional urban areas in Britain. Thiemann et. al. \cite{thiemann2010structure} applied two urban mobility techniques to investigate the geographical edges that emerge from an analysis of complex networks. Tannier et al. \cite{tannier2011fractal,tannier2013defining} integrated morphological approaches widely used in landscape ecology based on the analysis of the heterogeneity of the urban-rural interface for boundary identification. On the other hand, Masucci et al. \cite{masucci2015problem} and Arcaute et al. \cite{arcaute2016cities} used percolation analysis to find urban boundaries. They applied a clustering technique on the planar network of street-intersections and satellite images showing the strong dependence of scaling relationships on urban boundary definitions. Recently, Cottineau \& Vanhoof \cite{cottineau2019mobile} evaluated the prediction power of mobile phone datasets to understand demographic data on different urban extents based on population density. Finally, Humeres \& Samaniego \cite{humeres2017dissecting} proposed a spectral definition of cities based on the analysis of mobility networks using classical origin-destination (OD) surveys in Chile. This study builds on the methods developed by the last-mentioned work, broadening and strengthening its scope.

The growing availability of Information and Communication Technologies (ICTs) has opened new possibilities to contribute towards a more effective estimation of the urban extension limits. The high latency of recorded transactions in Call Detail Records (X/CDR) allows capturing the dynamic process of space usage among urbanites at high spatio-temporal and contextual resolutions  \cite{blondel2015survey}. Such unprecedented feature further permits to explore new solutions to the social problems of the urban organization \cite{blondel2015survey, gonzalez2008understanding, louail2014mobile, dannemann2018time}. This fundamental characteristic has provided an important ground to evaluate what Batty has termed \textit{The New Science of Cities} \cite {batty2013new}, a singular approach to understand the urban organization that places the network of interactions between people and urban infrastructure as the central aspect shaping the city. Studies involving ICTs have thus emerged as an essential tool to delineate the city, often geared towards a near real-time functional representation that goes beyond the usual limitations of comprehensive censuses, campaigns and static surveys to represent the urban environment.

This work proposes a data-driven approach allowing to distinguish urban (``U'') from non-urban spaces (``NU'') by analyzing the network of spaces built from the trajectories of urbanites. This methodology was applied to all cities with more than $100,000$ inhabitants in Chile. We used a centrality index as a proxy of the degree of concurrency throughout the city. To untangle NU spaces from U, first, each exploration area was partitioned, and second, the relationship between urbanites and the number of calls made in each U and NU partition was analyzed. This, ultimately allowed us to use the significant differences in scaling between social voice-calls and the number of urbanites (i.e. $ \beta_U $ and $ \beta_{NU} $) to distinguish and define urban spaces. 

\section{Materials \& Methods}
\label{S:2}

\subsection{Dataset}
The data used in this work corresponds to mobile telephone records generated by pre- and post-payment users to one of the largest provider in Chile, Telef\'onica ($\sim 37$ \% of market share). We used two datasets: (1) internet connection detail records (XDR) generated when using mobile internet and (2) voice call detail records (CDR). A total of 4 weeks of data were analyzed in March, May, October and November of 2015 with a total of $\sim 5.6 \times 10^6$ users throughout Chile and $\sim 3 \times 10^9 $ XDR records. Each record contains the date and time of connection, an anonymous identifier for each user and the geolocation of the antenna to which the connection is established. About $ 1 \times 10^9 $ CDR records were used to assign the number of calls registered to each antenna during the study period. 

The coverage area of antennas was approximated using Voronoi cells. This allowed to use XDRs as a proxy of individual mobility, since they represent the set of physical locations effectively visited by each user in his or her daily trajectory. CDRs, on the other hand, are proxies of the level of social voice interaction between urbanites; these were used as indicators of effective social cohesion and to define urban boundaries through urban scaling analysis. 

\subsection{Defining local interactions}
We took advantage of the high temporal density of XDR data and ascribed the residence of each urbanite to a particular location to only consider local mobility and communication in the analyses per city. To ensure a correct assignation of the home location for every user, we followed previous work done by our research group and others \cite{dannemann2018time, phithakkitnukoon2012socio, vscepanovic2015mobile, vanhoof2018detecting}. In brief, correct assignation of home location was possible by identifying the most frequented tower between 10 PM and 7 AM on weekdays with at least five pings. We additionally required at least a quarter of total pings for each user to be made within his or her residence during the four weeks dataset, thereby minimizing uncertainties associated with seasonal effects (e.g. changing movement behaviour of people going on holidays) among others.

\begin{figure}[h]
  \centering
  \includegraphics[width=.9\textwidth]{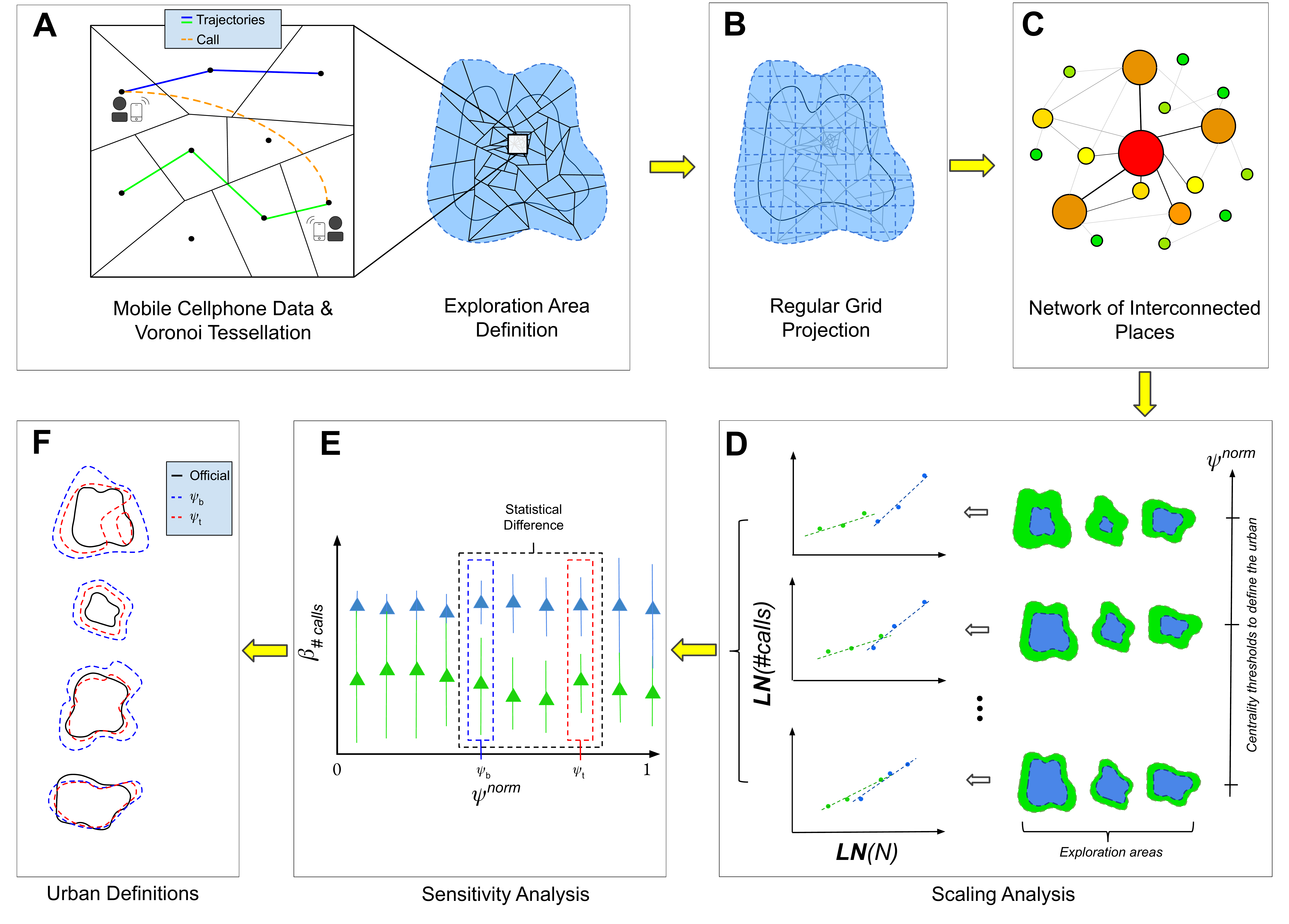}
  \caption{ Schematic representation of the analysis procedure. (\textbf{A}) X/CDR datasets were used to compute user's trajectories as a sorted list of visited places across exploration areas for each city. (\textbf{B}) A $1km^2$ grid was used as spatial analysis unit. (\textbf{C}) A network of interconnected places was created for each exploration area and centrality ranking ($\psi$) was computed for each node. (\textbf{D}) Thresholds were defined based on $\psi$ and the slope between the number of calls and the network size within and outside boundaries was evaluated. (\textbf{E}) Differences between slopes were evaluated in order to select the top and bottom values of the interval where urban and non-urban places are statistically different given the scaling relationship. (\textbf{F}) Comparison of resulting boundaries and official definitions of cities.
    \label{f:infog}
  }
\end{figure}

\subsection{Urbanite's trajectories and exploration areas}

Trajectories for each resident were constructed as a sorted list of visited locations across the day starting from his or her home location (Fig. \ref{f:infog}A). To estimate buffer sizes, the total distance for each trajectory was calculated as the sum of the geographic distance between visited locations and these were later used to frame the individual analyses within the exploration areas for each city (see below). 

In addition, to restrict the analysis to urbanites residing in each city, we only considered cities larger than 100,000 inhabitants in 2015 \cite{minvu}. This allowed to avoid the potential bias of having an unequal market share among telephone providers in smaller cities (i.e. population $< 10^5$). Official limits for the 21 selected cities were used as a starting area of exploration to evaluate the spatial extent of each city. A buffer was built around each city extending the official boundary (i.e. defined by the Ministry of Housing and Urbanism, MINVU) to the median of the total traveled distance of urbanites within the city (Fig. \ref{f:infog}B).

\subsection{Spatial partition of exploration area}
Voronoi cells were partitioned into a regular grid to minimize biases associated with the irregular distribution of antennas across cities. While several grid sizes could be considered, a $1 \times 1 km$ grid was just under the median area of Voronoi cells across the officially defined urban system analyzed (Fig. S1), and was hence chosen as a compromise that would provide a meaningful geographic analysis scale commensurate to official, vector-based outlines across cities (Fig. \ref{f:infog}B). This offered a granularity that further allowed a homogeneous visualization of the spatial distribution of antennas, as proposed in Louail et al. \cite{louail2014mobile}.

\subsection{Building networks of interconnected locations}
\begin{figure}[h]
  \centering
  \includegraphics[scale=0.4]{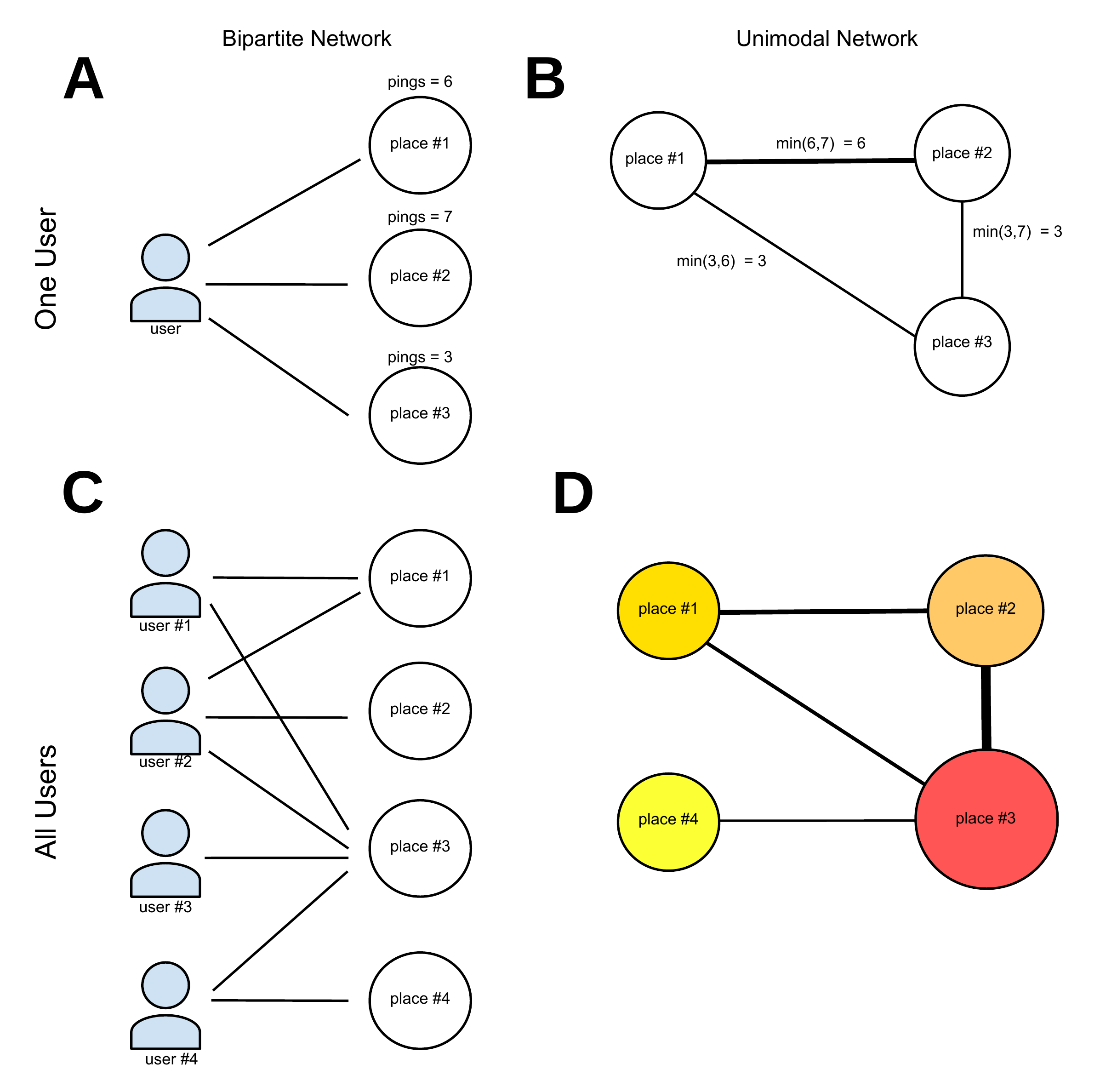}
  \caption{
    Unimodal projections of bipartite graphs. (\textbf{A}) Simple bipartite network for a single user visiting three locations; (\textbf{B}) Unimodal projection of locations visited by user in (A); (\textbf{C}) Bipartite graph representing places visited by four users; (\textbf{D}) Unimodal representation of (C). Centrality ranking is shown in red and the weight of the relationship between places is shown by the thickness of links between places. Similarity between two connected locations corresponds to the smallest amount of \textit{pings} between them. Each link is sequentially added to the user's trajectory increasing the weight $\omega_{i,j}$ of the link (Eq. \ref{E:min-pings}).
  }
  \label{f:network}
\end{figure}

Each urban system was represented as a network (Fig. \ref{f:infog}C). This sought to depict the interaction between individuals and the urban infrastructure. For this, a bipartite graph was used to represent the set of locations visited by each user on a daily basis within the exploration area (Fig. \ref{f:network}). Subsequently, the bipartite graph was transformed into a weighted unimodal graph of interconnected locations (Fig. \ref{f:network}B \& D). Such a graph allowed us to analyze the network of locations (i.e. antennas as nodes) visited by each person (i.e., links). Interaction rates between locations, $\omega_{i,j}$, were computed as the sum of the smallest number of pings (i.e., XDR records), $p$, between locations (i.e. $i$ and $j$) visited by user $u$. Note that this is, in fact, a similarity measure equivalent to computing the intersection of pings among the total number of pings per user, $p_{total}$; see Eq. \ref{E:min-pings} \cite{humeres2017dissecting, humeres2014power}.

\begin{equation}\label{E:min-pings}
    \omega_{i,j} = \sum_{u} \frac{min(p_{i}^u, p_{j}^u)}{p_{total}^u} 
\end{equation}
We normalized the interaction rate between locations, $\omega_{i,j}$, over the sum of pings, $p_{total}$, for each user in order to compare weights across locations.

\subsection{Spectral analysis of locations networks}
Eigenvector centrality was used to estimate the importance of each cell in the city. It characterized nodes (i.e. cells) in the network as a function of the number of visits connecting these locations. Interestingly, centrality values are independent of the order in which they are visited \cite{borgatti2005centrality}. The Bonacich method was used to compute centrality \cite{bonacich1972factoring}. Each cell was then ranked, so that better-connected cells (i.e. linked to many places) would have higher values of centrality and thus a higher ranking. 

Eigenvector centrality ranking is defined as $ \psi_{i} = u_{1, i} $, where $u_{1,i}$ is a component of $u_1$, the eigenvector associated with the largest eigenvalue $\lambda_1$  satisfying $Au_1 = \lambda_1 u_1$  \cite{newman2006modularity}. Centrality calculation was performed using a power iteration method, as implemented in graph-tool v2.27 \cite{peixoto_graph-tool_2014, langville2005survey}.

A discrete, 0.01, range-interval of centrality values was defined to compare computed centralities across cities ($\psi^{norm}$). Therefore, the kurtosis of $\psi$ suggested the following logarithmic transformation:

\begin{equation}\label{E:normrank}
	\psi^{norm}_{i,x} = \frac{\log \psi_{i,x}-min(\log \psi_x)}{max(\log \psi_x)-min(\log \psi_x)} ;\ 
    \psi^{norm}_{i,x} \in [0,1].
\end{equation}

\subsection{Urban scaling analysis} 
Urban scaling analysis was used to evaluate any potential regime change across urbanites' social voice-interactions for the various spatial extents defined by a threshold in $ \psi^{norm}$ (Fig. \ref{f:infog}D). The elasticity, β, of the scaling relationship between population $N(t)$ and the number of voice-calls $Y(t)$ (see below), was then compared for significant differences (Fig. \ref{f:infog}E). Y (t) was measured from an independent CDR dataset for the same time period. The urban scaling model considered the size of the city $ N(t) $, measured by the number of urbanites making voice calls at time t in a defined area \cite{bettencourt2007growth}:

\begin{equation}\label{E:urbanscaling}
    Y(t) = Y_0N(t)^\beta
\end{equation}

$Y (t)$ has often been associated to measurable urban characteristics (e.g. income, social interaction, extension of electrical wiring, etc.), $Y_0$ is a normalization constant and the elasticity, $\beta$, characterizes the general dynamics and growth regime across a gradient of city sizes \cite{bettencourt2007growth,bettencourt2010unified}. Equation \ref{E:urbanscaling} is parameterized using ordinary least squares method after a $log_{10}$ transform: $ log_{10} \,(Y (t)) = log_{10} \, (Y_0) + \beta \, log_{10} \, (N (t) ) $.

With the premise that urban boundaries are shaped by the interaction between people and the urban infrastructure, Eq. \ref{E:urbanscaling} was used as a diagnostic tool to identify the critical interval [$ \psi_b $, $ \psi_t $] that significantly --and functionally-- bounds the urban space. For this, we sequentially partitioned the exploration area (i.e. \textit{buffer} + conurbation MINVU) into two sets of locations. The first set of locations , had a centrality rank larger than the defined threshold value (e.g. $\psi_b$) and, the second, included locations not meeting such condition (Fig. \ref{f:infog}). The hypothesis of $\beta$'s belonging to the same distribution was evaluated using standard \textit{t-test} \cite{andrade2014statistical}. Places ranking equal or greater than the threshold were evaluated using Eq. \ref{E:urbanscaling} and labeled as urban (U), while those ranking lower were assigned to non-urban spaces (NU). This method allowed to spatially partition the exploration area in two different systems and compare $\beta_U$ and $\beta_{NU}$ elasticity. Finally, we selected the largest continuous ranking interval having statistically different $\beta$ (Fig. \ref{f:infog}E), generated the corresponding map of U, NU, and qualitatively compared our results to the official delineation of 21 cities in Chile.

\section{Results}
\label{S:3}
\begin{figure}[!h] 
  \centering
  \includegraphics[width=.7\textwidth]{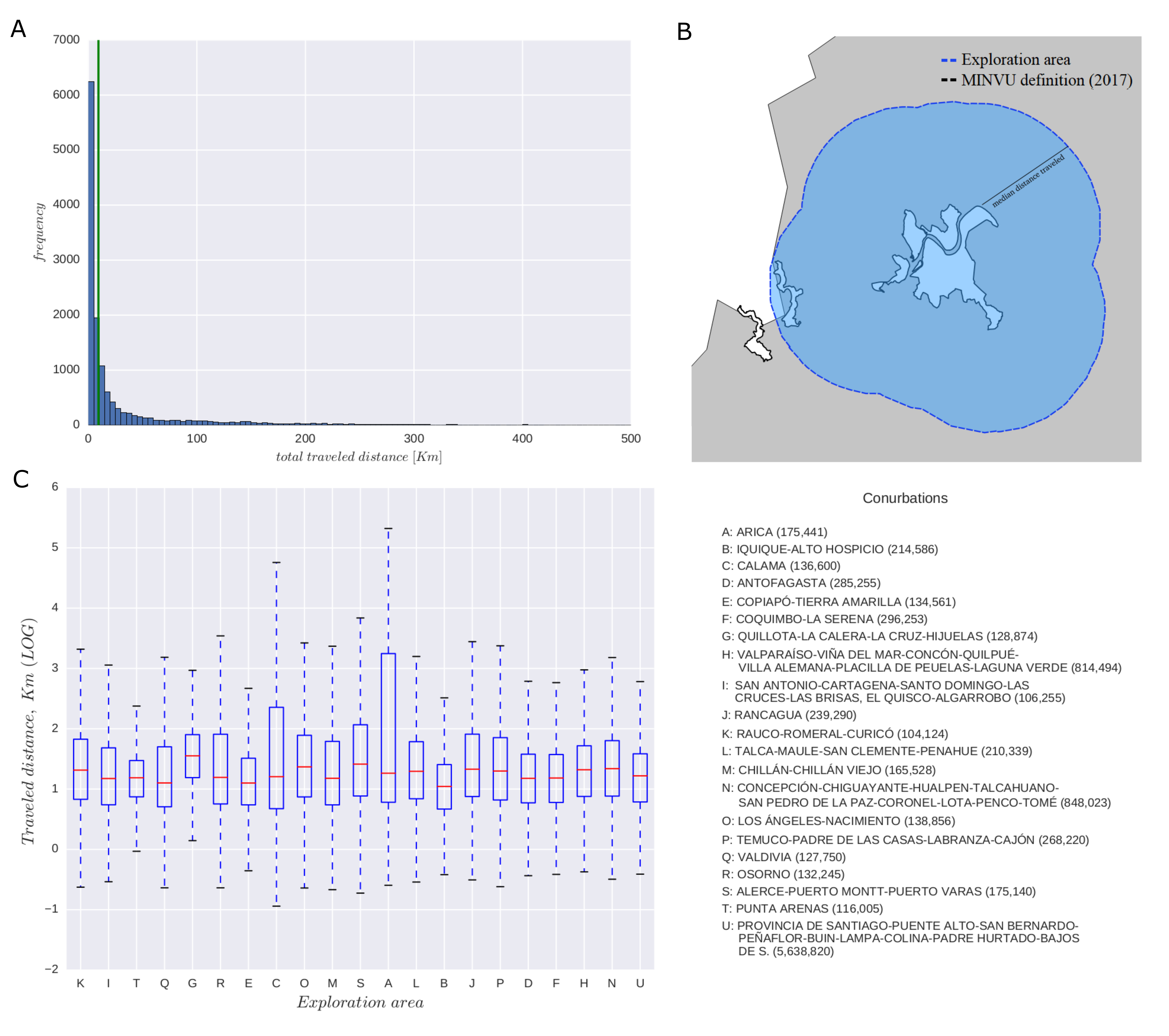}
  \caption{
    (\textbf{A}) Distribution of total distance traveled by mobile telephony users in Valdivia. The vertical green line shows the median traveled distance (9.31 Km) over all cities. (\textbf{B}) An example of the buffer creation process for the city of Valdivia. The official urban area for 2017 is shown in lighter blue within dark blue buffer area, as defined by the Chilean Ministry of Housing and Urbanism (MINVU) \cite{minvu}. (\textbf{C}) Box-plot of traveled distances within the buffer exploration area for each individual city considered in this study. Cities are ordered from the smallest on the left to the largest on the right. Population sizes are in parenthesis along with labels for (C). 
  }
  \label{f:buffer}
\end{figure}

Traveled distance among cities was found to be fairly constant and showed a right-skewed distribution (Fig. \ref{f:buffer}A) with an overall median of $\sim 9.3$km across exploration areas. Figure \ref{f:buffer}B shows an example of the exploration area for the city of Valdivia (Q).

\subsection{Eigen-vector ranking analysis}
The centrality ranking explicitly divided the exploration areas based on the overlap of human activities for each city. Not surprisingly, our analysis showed that central locations are within the official city limits with a decay of its centrality rank towards the periphery.
\begin{figure}[!h]
  \centering
  \includegraphics[width=.9\textwidth]{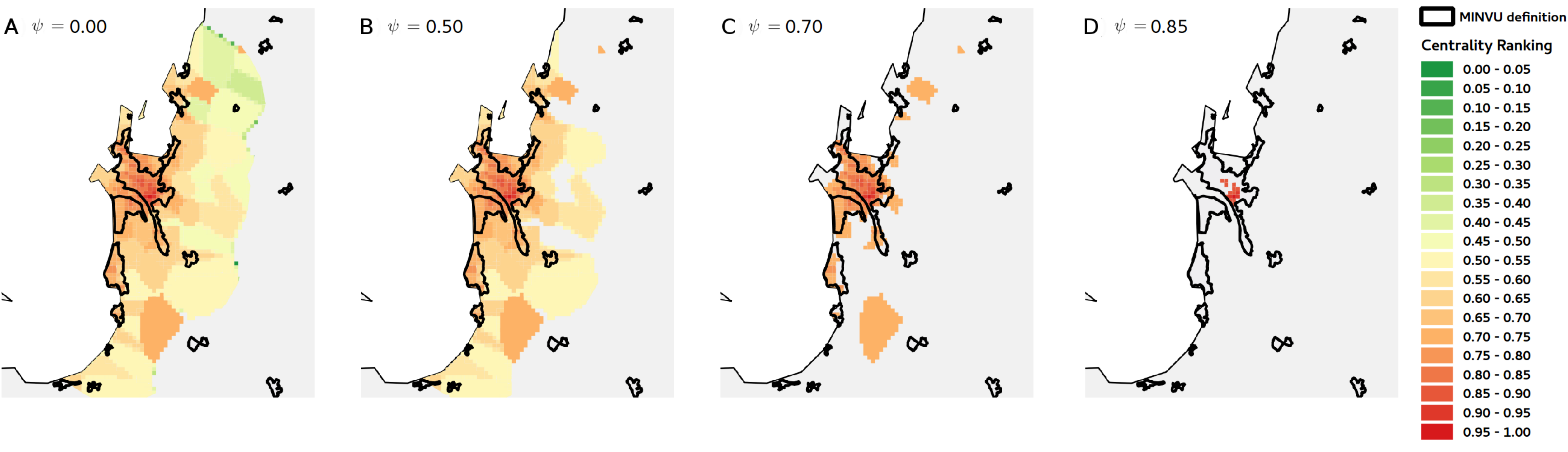}
  \caption{
    Spatial distribution of centrality ranking using four thresholds ($\psi^*$) across the exploration area for the city of Concepci\'on for Sunday, March 15, 2015. (\textbf{A}) Centrality within the exploration area ($\psi^* \geq 0.00$), (\textbf{B}) $ \psi^{norm} \geq 0.50 $, (\textbf{C}) $ \psi^{norm} \geq 0.70 $, (\textbf{D})  Locations with $ \psi^{norm} \geq 0.85 $. The base map was projected to S19 UTM coordinates and WGS84 datum.
  }
  \label{f:buffer-ranking}
\end{figure}
For instance, in the case of Concepci\'on, the third-largest city in Chile, central locations roughly coincided with official urban boundaries in spite of the intricate configuration of city limits. Moreover, the approach proposed here highlighted additional locations with medium centrality ranking, which were not considered within official city boundaries (Fig. \ref{f:buffer-ranking}).

\begin{figure}[!h]
  \centering
  \includegraphics[width=.9\textwidth]{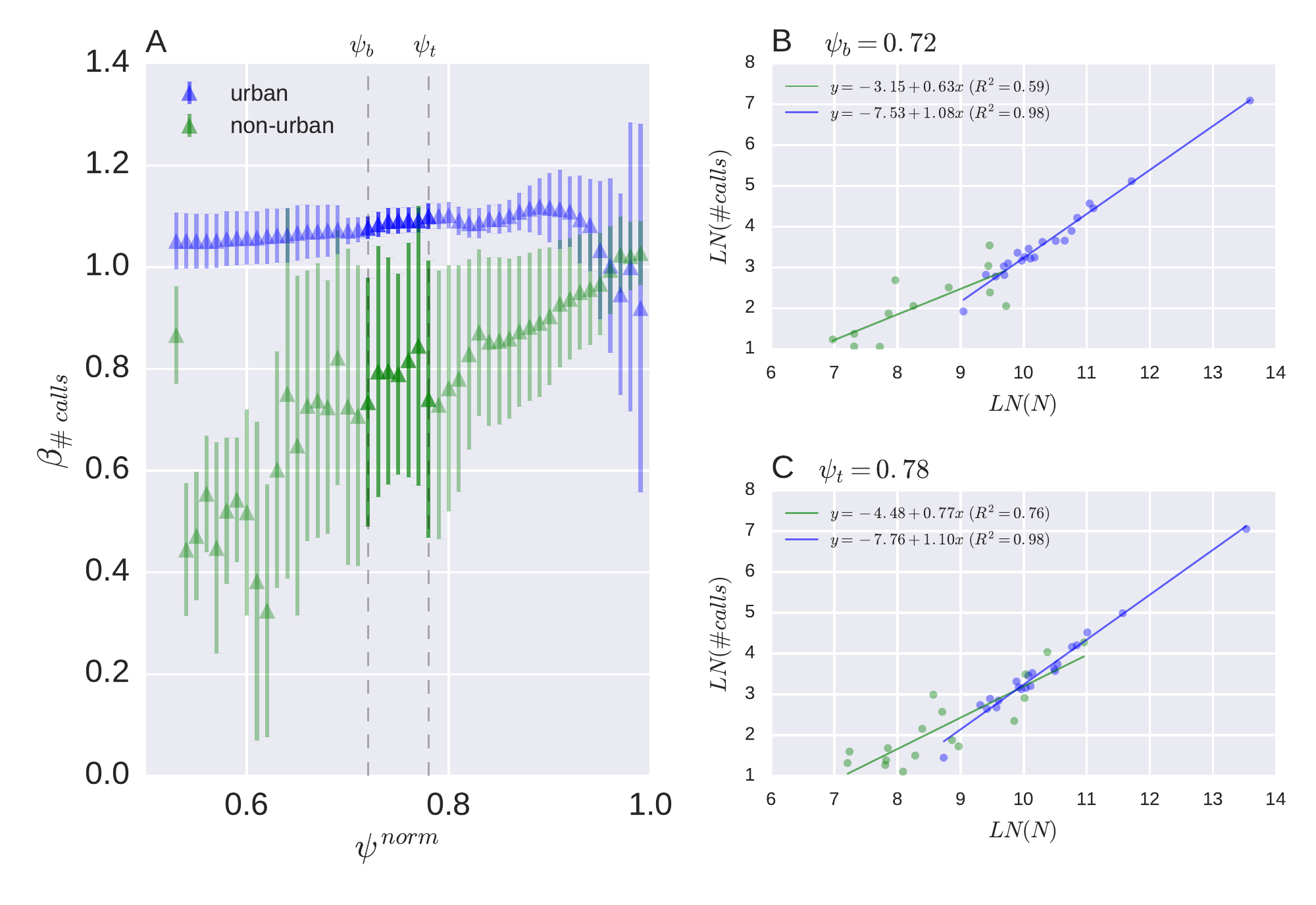}
  \caption{
    Sensitivity analysis of exponents $\beta_U$ and $\beta_{NU}$ for different thresholds of ranking ψnorm. Securities of exponents $\beta_U$ and $\beta_{NU}$ were obtained at 95\% confidence. If the slopes were not statistically different or if the errors on the y-axis were not normally distributed for both systems, their comparison is presented with transparency. The interval obtained where both systems are distinguished corresponds to [$0.72$, $0.78$] for the day Sunday, March 15$^{th}$, 2015.
  }
  \label{f:scaling-law-calls}
\end{figure}

The larger elasticity recorded for U compared to areas labeled as NU, suggests that urban scaling analysis is an appropriate diagnostic tool to differentiate visitation regimes across the boundary delineation. In fact, $\beta_U$ was clearly larger than βNU over the full range of ψnorm, with standard errors for $\beta$'s larger at the margins of $\psi^{norm}$. This was most likely due to the lower number of cells labeled as U at lower $\psi^{norm}$ and, conversely, a smaller number of cells labeled as NU where large centrality usually prevailed (Fig. \ref{f:scaling-law-calls}). In spite of this, it was possible to identify a centrality interval between $[0.72,0.77]$ that denoted statistical differences in voice-call interactions between U and NU places after evaluating the hypothesis of $\beta$ being drawn from the same population (detailed sensitivity analysis available in Supplementary Material).

\begin{figure}[!h]
  \centering
  \includegraphics[width=.7\textwidth]{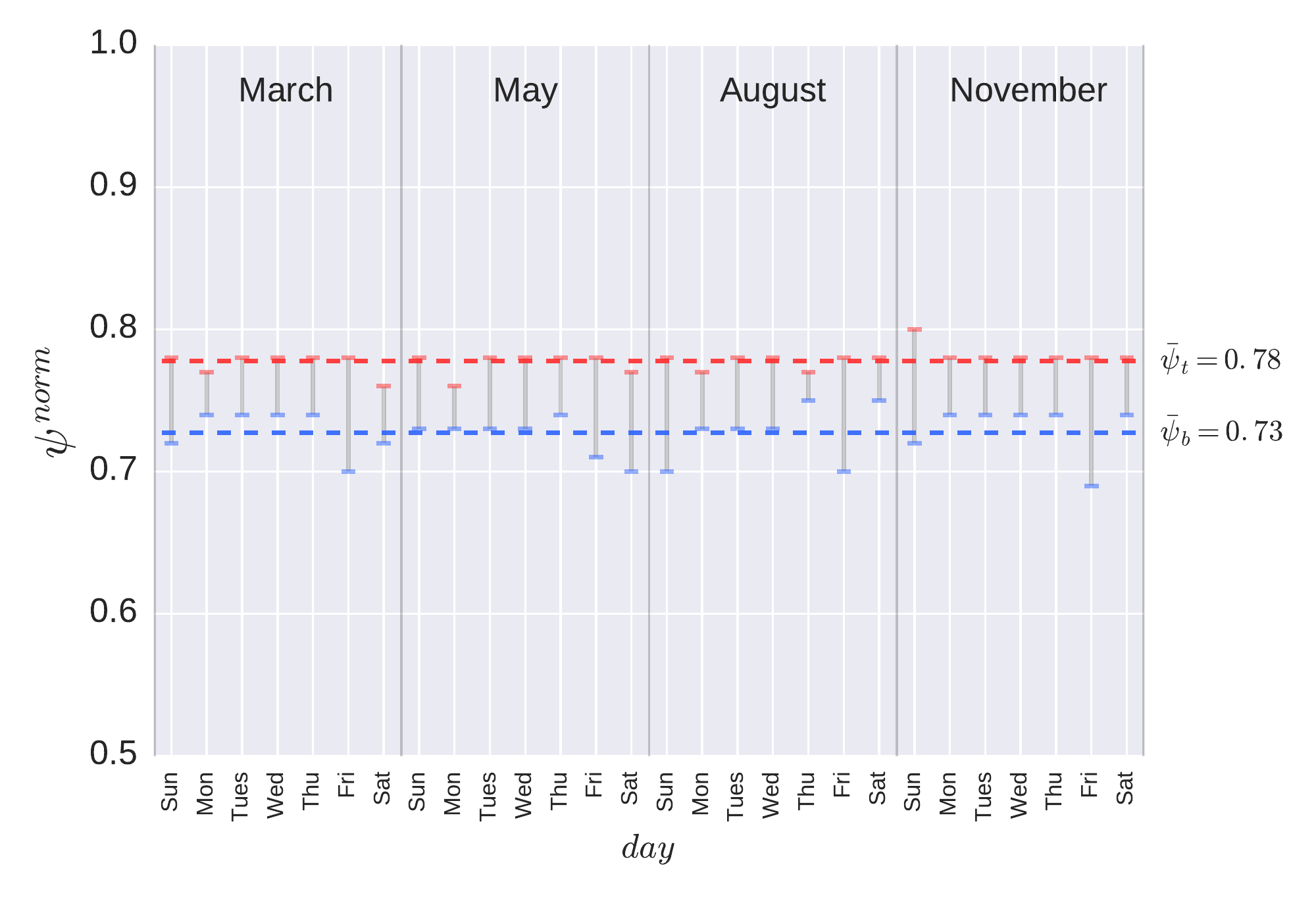}
  \caption{
    Change in the definition intervals of the cities for the days analyzed. $ \bar\psi_{b} $ (blue) corresponds to the average of the lower limit and $ \bar\psi_{t}$ (red) to the upper limit of the intervals where the city is defined for the different days analyzed.
} 
  \label{f:def-by-time}
\end{figure}

A complete analysis for each day in our records showed a stable distribution of centrality intervals (Fig. \ref{f:def-by-time}). The values of $ \psi_b $ and $ \psi_t $ were bounded in the range [$0.7$, $0.8$] with lower and upper mean values of $ \bar{\psi_{b}} = 0.72 $ and $ \bar{\psi_{t}} = 0.78 $, respectively.

\subsection{Eigen-cities: Cities boundaries from urban mobility and eigen-vector centrality of places}

\begin{figure}[!h]
  \centering
  \includegraphics[width=.7\textwidth]{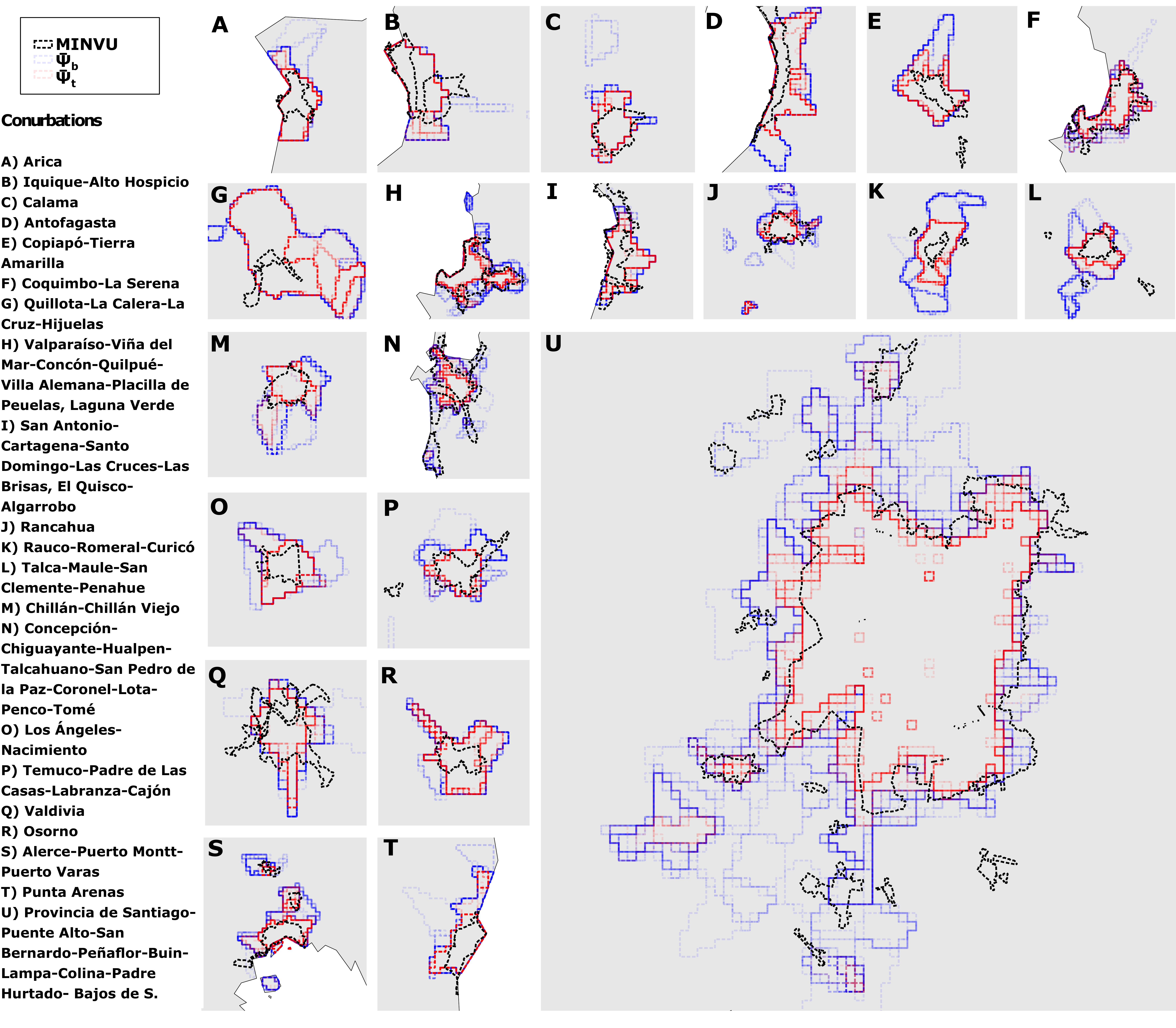}
  \caption{
    City limit comparison across the full dataset. The upper right box shows the different threshold definitions. Official limits are in black \cite{minvu}, $ \bar {\psi_{b}} $ threshold in blue, and $ \bar {\psi_{t}} $  in red contour. To visually show the most recurrent limit throughout the dataset, limits for each day are plotted using transparency so that more frequent limits become apparent due to the spatial superposition of city limits. 
  }
  \label{f:city-boundaries}
\end{figure}

Spatial comparison of urban extents based on the upper centrality threshold ($ \bar{\psi_{t}} $) showed extensive morphological similarities with official city limits  \cite{minvu} (Fig. \ref{f:city-boundaries}). The complex nature of the spatial comparison made the quantitative comparison a complex task. Hence, we provided a qualitative analysis in which we overlaid city limits computed for each day using transparent layers (Fig. \ref{f:city-boundaries}). This procedure made explicit the more recurrent limits across the dataset, thereby providing a better outlook of the differences between the official city limits and those proposed by our method. An additional, and more detailed, analysis procedure is provided in the Supplementary Materials (Fig. S3) along with quantitative areal differences for the cities of Concepci\'on, Valpara\'iso, and Santiago (Table S1).

Some differences could be observed at the peripheries (i.e., (H) Vi\~na-Valpara\'iso, (F) La Serena-Coquimbo, and (U) Gran Santiago). Additionally, cities with low density of antennas seemed to show a tendency to depict larger urban areas compared to the official delimitation. This is most likely due to biases associated with the process of moving from Voronoi to the 1km$^2$ grid cells. Since antennas are originally placed to ensure connectivity and not to uniformly cover the territory, a sparser antenna distribution occurs towards more rural areas. Therefore, our method can potentially identify functional settlements detached from the central core of the study area, such as the cases of (J) Rancagua, (N) Gran Concepci\'on, (S) Alerce-Puerto Montt-Puerto Varas, and (T) Punta Arenas (Fig. \ref{f:city-boundaries}).

\section{Discussion}
This paper presented a methodology to delimit the urban extent with a functional perspective and using large volumes of non-conventional data. We used spectral analysis to unveil the centrality of places as nodes in the network of commonly visited places. Urban boundaries were evaluated through the rate of voice-call interactions within an exploration area. We applied our method to a concrete example delineating major urban systems in Chile, explicitly showing the advantages and opportunities offered by ICTs.

As an alternative to proposals based on the analysis of population density, distance thresholds, and extensive surveys \cite{rozenfeld2008laws, oliveira2018worldwide, humeres2017dissecting}, we combined the dynamics of place visitations with voice interaction to define urban boundaries. Some of the previous studies have focused on studying the fractal geometry of urban textures to define the extent of cities \cite{tannier2011fractal, tannier2013defining, batty1994fractal, batty1996preliminary, frankhauser1998fractal}, while others have contributed with the analysis of topological patterns of street networks \cite{masucci2015problem, jiang2007topological} and the delineation of socially cohesive areas through complex networks analyses and various types of thresholds \cite{ratti2010redrawing, grady2012modularity}.

Important contributions have already shown the importance of using ICT for understanding the structure and dynamics of cities \cite{Cranshaw2012}. Cottineau \& Vanhoof \cite{cottineau2019mobile} have recently addressed the limitations of big-data in the generation of traditional urban indicators. Using arbitrary definitions of cities they evaluated how measures drawn from CDR may statistically relate to socioeconomic indicators. They showed a compelling example that highlighted the importance of having an adequate, and functional definition of the urban boundaries, as it was proposed in their earlier work \cite{cottineau2018defining}. On the other hand, we improved methods used by previous contributions by proposing the use of an urban boundary definition based on empirical observations of social interactions measured through voice-calls between individuals. We explicitly showed how large volumes of urbanites' trajectories may be processed to obtain near real-time urban delineation. In fact, while we used CDRs, it is worth noting that other spatially explicit indicators may also be used as a diagnostic tool to measure differences between urban and non-urban systems as done by \cite{Cranshaw2012}.
Nowadays, given the increasing availability of big datasets makes our approach even more relevant. Our method allows to explore temporal constraints or the variation of urban boundaries across the day, as proposed by the Time Geography research program \cite{Hagerstrand1970,Justen2013}. Using empirically-based descriptions of the heterogeneous usage of infrastructure, this method can aid in finding new planning perspectives to accommodate human movement (or existing/new infrastructure) throughout the city. Additionally, it may even prove instrumental for implementing participatory methodologies of urban informatics \cite{FOTH2018}.

The simple heuristic method proposed here can clearly be improved. In fact, building a network of places by considering the minimum number of pings between locations in order to weigh their relationship (i.e. $\omega_{i,j}$) is only a starting point to capture the overlap of social interactions. Several other approximations may be possible to account for the way in which urbanites connect places. For instance, we might be interested in considering the average number of visits between two locations as a more representative indicator of the relationship between places.

Ranking places by their centrality value allows identifying key locations deserving special attention in terms of urban planning, transport, social studies and marketing, among others. An early proposal by Zhong et. al. \cite{zhong2017revealing} designed to describe and understand polycentric cities, defined central spaces according to the number and diversity of activities. Our approach may provide a technical framework to extend this methodology and identify not only the spatial extent of cities but also depicting locations with meaningful urban activity for planning purposes. Our method may additionally offer the opportunity to use near real-time data to evaluate the state of cities and exposing a series of questions relevant to urban planning. For instance, could this dynamic delineation of cities help to solve specific transportation issues? Can we plan ahead of time when organizing the urban infrastructures during particular events? We are confident that our method could provide such planning tools. However, further research is still needed as, for example, we need to develop new computational tools capable of modeling trajectories across the urban environment.

We believe that this paper offers a data-driven methodology incorporating the notion of cities as dynamic entities defined by individuals and their daily interactions, as proposed by the New Science of Cities \cite{batty2013new}.

\section*{Funding}
Funding for this research was provided by CONICYT under the grant FONDECYT \# 1161280 to HS.

\bibliographystyle{elsarticle-num}

\bibliography{Sotomayor_Samaniego.bib}

\section{Supplementary Material}


\end{document}